# Picosecond spin-orbit torque switching of ferrimagnets

Hao Wu[1,*], Deniz Turan[1], Quanjun Pan[1], Guanjie Wu[2], Seyed Armin Razavi[1], Nezih Tolga Yardimci[1], Zongzhi Zhang[2], Mona Jarrahi[1], and Kang L. Wang[1,†]

[1]*Department of Electrical and Computer Engineering, University of California, Los Angeles, California 90095, USA*

[2]*Shanghai Engineering Research Center of Ultra-Precision Optical Manufacturing, Department of Optical Science and Engineering, Fudan University, Shanghai 200433, China*

[*]Corresponding author: wuhaophysics@ucla.edu

[†]Corresponding author: wang@ee.ucla.edu

**Abstract:**

Spintronics provides an efficient platform for realizing non-volatile memory and logic devices. In these systems, data is stored in the magnetization of magnetic materials, and magnetization is switched in the writing process. In conventional spintronic devices, ferromagnetic materials are used which have a magnetization dynamics timescale of around the nanoseconds, setting a limit for the switching speed. Increasing the magnetization switching speed has been one of the challenges in spintronic research. In this work we take advantage of the ultrafast magnetization dynamics in ferrimagnetic materials instead of ferromagnets, and we use femtosecond laser pulses and a photoconductive Auston switch to create picosecond current pulses for switching the ferrimagnet. By anomalous Hall and magneto-optic Kerr (MOKE) measurement, we demonstrate the robust picosecond SOT driven magnetization switching of ferrimagnetic GdFeCo. The time-resolved MOKE shows more than 50 GHz magnetic resonance frequency of GdFeCo, indicating faster than 20 ps spin dynamics and tens of picosecond SOT switching speed. Our work provides a promising route to realize picosecond operation speed for non-volatile magnetic memory and logic applications.

**Main Text:**

Electrically manipulation of the magnetization is both fundamentally interesting and practically important for spintronic devices[1,2]. To date, two crucial issues need to be addressed: (1) reducing the energy consumption[3] and (2) increasing the switching speed. There have been several approaches to reduce the energy consumption, such as voltage-controlled magnetization[4,5], the giant spin-orbit torque in topological insulators[6-11], and so on. As for the switching speed, up to now, most spintronic devices are still operated in several nanoseconds' timescale, which corresponds to the GHz magnetic resonance frequency of ferromagnetic materials. Ferrimagnets have antiferromagnetically-coupled spin sublattices, as a result, the exchange coupling-induced magnetic resonance frequency is on the order of 100 GHz near their compensation point[12], which can enable picosecond magnetization switching speed.

There have been efforts for realizing ultrafast all-optical switching of ferrimagnetic materials[13], with circularly-polarized femtosecond laser pulses resulting in switching speeds on the order of tens of picoseconds[14-17]. Similarly, laser-induced hot-electron pulses have been exploited to induce ultrafast switching dynamics in ferrimagnetic materials[18-20]. However, these methods mainly rely on heating effects and have relatively low energy efficiencies compared to electrical methods. Therefore, picosecond electrical switching of ferrimagnets provides a promising route towards ultrafast spintronic devices with high energy-efficiency, which has not been experimentally explored yet.

In this work, a photoconductive Auston switch is employed to generate the picosecond electrical pulse from a femtosecond laser pulse[21], which is used to provide the picosecond spin-orbit torque (SOT) switching[22-24] in Ta/GdFeCo heterostructures. The magnetization switching is detected by both the anomalous Hall effect (AHE) and the magneto-optic optical Kerr effect (MOKE), which demonstrate robust magnetization switching by the picosecond SOT with a peak current density of $10^7$ A/cm$^2$. Furthermore, 50 GHz spin dynamics of GdFeCo is detected by the time-resolved MOKE (TR-MOKE), further confirming the tens of picosecond switching speed of ferrimagnets.

Figure 1a shows the schematic of the device structure and the measurement method, where a photoconductive Auston switch is connected to the magnetic strip of Ta/GdFeCo heterostructures. When the femtosecond laser pulse is focused on the gap of the Auston switch, the charge carriers will be excited in the conduction band of low temperature grown GaAs (LT-GaAs), which are subsequently accelerated by the bias voltage. Based upon simulations in the Sentaurus software packages, the induced photocurrent response time is estimated to be around 1 ps. The magnetization of GdFeCo is detected either by the AHE or the MOKE, where an in-plane magnetic field is applied to break the inversion symmetry between $\pm M$ during the SOT switching. Figure 1b shows the microscopic picture of the Auston switch+magnetic strip device, where the width of the magnetic strip is 20 μm. The $R_{xy}$-$H_z$ curve in Figure 1c demonstrates the strong perpendicular magnetic anisotropy (PMA) of GdFeCo.

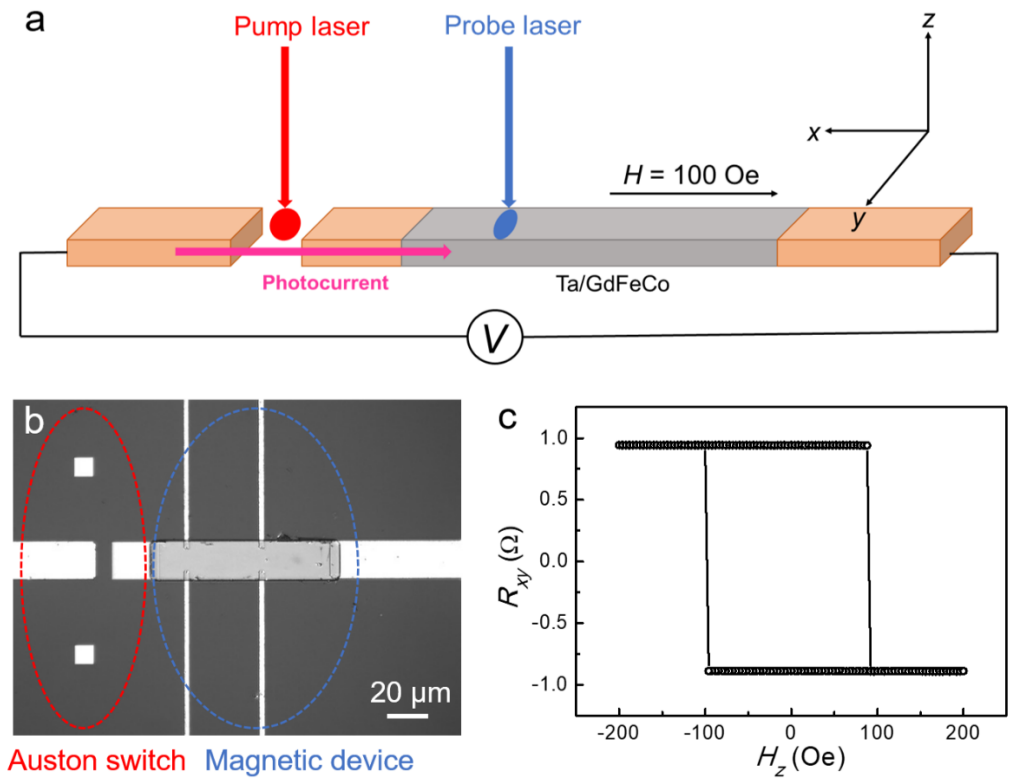

**FIG. 1.** a, Schematic of the Auston switch+magnetic strip device and the measurement set up. The THz laser pulse is focused on the gap of the voltage-biased photoconductive Auston switch to excite the picosecond current, which then provides a picosecond spin-orbit torque (SOT) on the ferrimagnetic GdFeCo. b, SEM picture of the patterned Auston switch+magnetic strip device. c, $R_{xy}$-$H_z$ curve shows the strong perpendicular magnetic anisotropy (PMA) of GdFeCo.

To test the device, we first perform all-electrical SOT switching of the Ta/GdFeCo heterostructure using relatively long electrical pulses. Figure 2a,b show the current-driven SOT switching, where the magnetization is detected by the AHE resistance. In this measurement scheme, a 1-ms writing current pulse is applied to the switch the magnetization firstly, and is followed one second later by another 1-ms reading current pulse to detect the AHE resistance. Under ±50 Oe in-plane magnetic field $H_x$, the current-induced switching polarity changes, which is the typical SOT characteristic. The switching current density $J_c$ is around $1.1 \times 10^7$ A/cm² for the 1-ms writing pulse, which is similar to the previous report[25]. MOKE microscope is employed to detect the magnetic domain directly, which also clearly show the robust SOT-driven magnetic domain switching, as shown in Figure 2c. In order to figure out the current-induced Joule heating effect on the SOT switching, we change the writing pulse width from 0.5 s to 10 ns, as shown in Figure 2d. $J_c$ significantly increases for the writing pulse width from 0.5 s to 1 ms, and then slightly changes from 1 ms to 10 ns, indicating Joule heating is not a dominant effect with shorter writing pulses.

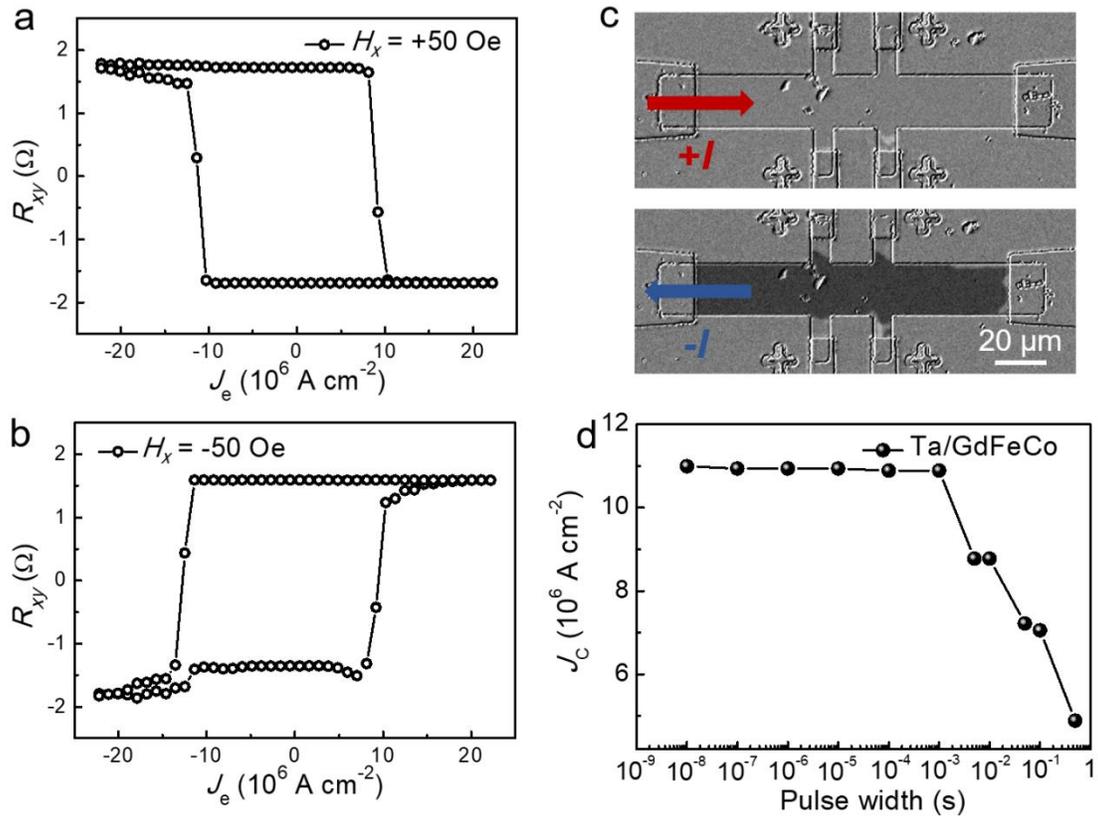

**FIG. 2.** a and b show the current-driven SOT switching of the Ta/GdFeCo heterostructure under $H_x = \pm 50$ Oe. c shows the SOT-driven magnetic domain switching measured by the MOKE microscope. d, Switching current density $J_c$ as a function of the pulse width of the writing current.

The magnetic precession period of the GdFeCo samples is more than 3 orders of magnitude shorter than the duration of the writing pulses in Figure 2d. It is therefore reasonable to expect that the SOT switching of the GdFeCo could be achieved in devices subjected to significantly shorter writing pulses. Previous time resolved magnetization studies of GdFeCo structures report full magnetization switching in structures subjected to ps and sub-ps length optical excitation. However, these previous results are both rooted in thermal effects, therefore, the demonstration of deterministic, ultrafast, electrical switching of GdFeCo magnetization is highly desirable from a technological perspective.

To test the temporal limitations of SOT switching in the Ta/GdFeCo heterostructure, we perform the experiments on devices subjected to ps duration electrical pulses. These ultrashort electrical excitations are generated by illuminating a DC biased photoconductive switch with the fs output of a mode locked Ti:Sapphire laser. A typical device used for these experiments is depicted in Figure 1b. Given the duration of the laser pulse (~150 fs) and the known carrier lifetime within the switch (~300 fs), electrical pulses on the order of 1 ps duration are predicted by simulations in the Sentaurius software package. By tuning the applied laser power (100 mW) and the bias voltage (10 V), average currents as high as 10 µA are observed in these devices, equating to a peak density of $10^7$ A/cm$^2$ for the picosecond current. When testing the viability of ultrafast SOT switching in our devices, we apply an in-plane magnetic field to the device under test to break the inversion symmetry between $\pm M$. Experiments are performed by illuminating the switch for approximately a minute, and then reading out the magnetization state through measurements of the AHE resistance (Figure 3a) and the Kerr rotation angle (Figure 3b). Significantly, we observe that the magnetization states of device are indeed changed when subject to this procedure. Unlike the previous report of ultrafast electrical switching in GdFeCo, however, we observe that the final magnetization state of our devices is dependent upon the polarity of the bias voltage across the switch as seen in Figure 3a,b. That the switching is dependent upon bias polarity indicates that this magnetization switching primarily comes from picosecond current-induced SOT, rather than as a consequence of thermal effects.

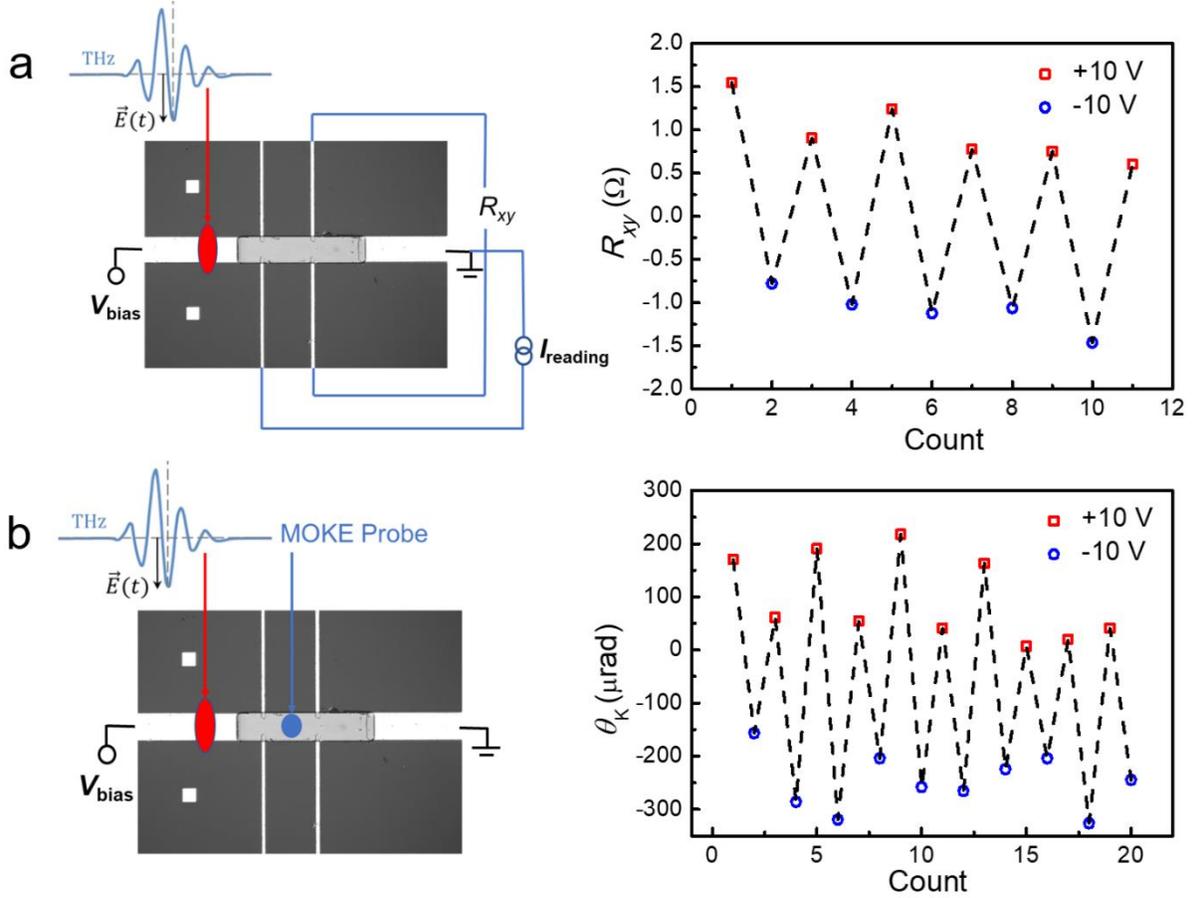

**FIG. 3.** a, THz laser pulse is applied at the gap the photoconductive Auston switch to excite the picosecond current under the bias voltage, and then provides a SOT on the Ta/GdFeCo heterostructure, where the magnetization is detected by the AHE resistance. When the bias voltage polarity is switched from +10 V to -10 V, the sign change of the AHE resistance demonstrates the magnetization switching from up to down states, which indicates SOT switching rather than thermal-induced switching. b, Picosecond-SOT driven magnetization switching detected by the MOKE.

While deterministic switching is observed in our ultrafast measurements, the measured Hall and Kerr signals provided in Figure 3a,b do not fully return to their initial state upon magnetization reversal, and appear to drift slightly over many switching cycles. The failure to recover the Hall resistivity or Kerr rotation observed at full magnetization is likely an indicator of only partial SOT switching in these devices, resulting in a multi-domain final state. That the Hall resistance and Kerr rotation of the final state drift following several switching cycles stems from an asymmetry in the current generated in our Auston switch upon the reversal of the external bias.

We have shown that the picosecond SOT can efficiently switch the ferrimagnetic GdFeCo, while another important question is how fast of the magnetization switching speed? In order to answer this question, we perform the TR-MOKE measurement[26-28] to pick up the spin dynamics of the ferrimagnetic GdFeCo. Figure 4a shows the typical TR-MOKE curve of the Ta/GdFeCo heterostructure under $H$ = 15 kOe. To extract precession frequency $f$, the dynamic Kerr signals are fitted by using the following expression[29]:

$$\theta_k = c_0 + c_1 \exp(t/t_0) + c_2 \sin(2\pi f t + \varphi)\exp(-t/\tau) \quad (1)$$

Here $c_0$ is the background signal. The second term is an exponential decaying signal representing the slow recovery process, where $c_1$ is the amplitude and $t_0$ is the characteristic relaxation time. $c_2$, $f$, $\varphi$, and $\tau$ in the third term represent the magnetization precession parameters of amplitude, frequency, initial phase, and decay time, respectively. The fitted frequencies under different $H$ are shown in Figure 4b, which is consistent with our expectations. From the fitting, we can get as large as 52.4 GHz magnetic resonance frequency of GdFeCo, which corresponds to a faster than 20 ps timescale of spin dynamics, and therefore, indicating the tens of picosecond speed of SOT switching.

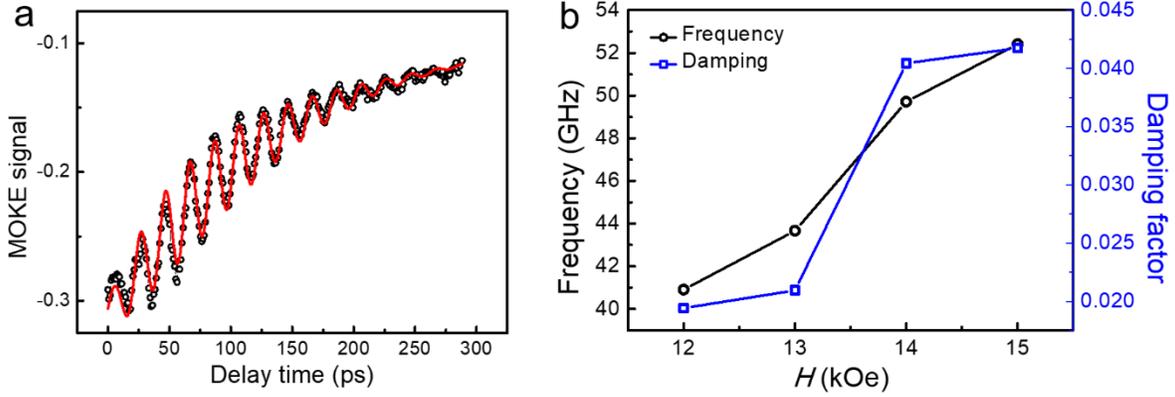

**FIG. 4.** a, Normalized dynamic TR-MOKE signals for the Ta/GdFeCo heterostructure. b, Magnetic resonance frequency and damping factor as a function of the magnetic field.

In summary, by combing the photoconductive Auston switch with the Ta/GdFeCo heterostructure, a picosecond writing current pulse is applied to provide the SOT, where both the AHE and MOKE measurements demonstrate the robust picosecond SOT-induced magnetization switching. The TR-MOKE measurement demonstrates faster than 20 ps spin dynamics, indicating tens of picosecond SOT switching speed of ferrimagnetic GdFeCo. These results demonstrate the picosecond SOT switching of ferrimagnets, which can increase the speed of spintronic devices from current ns to ps ranges.

**Experimental Section**

Picosecond electrical pulses are generated using a photoconductive Auston switch based on the low-temperature-grown gallium arsenide (LT-GaAs). LT-GaAs has a high mobility and a high intrinsic resistivity, and the photogenerated carriers exhibit a short lifetime (300 fs), which leads to the short photocurrent response time (1 ps). A mode locked Ti:Sapphire laser operating at 80 MHz is used to excite the Auston switch. In transport experiments, 800 nm pulses are focused on the gap of Auston switch using a microscope objective. After illuminating the Auston switch for approximately one minute, the beam is blocked, and the AHE resistance is read out using a four-contact measurement. All-optical experiments are conducted using the same protocol of minute-long illumination of the Auston switch, followed by a measurement of the Kerr rotation at the magnetic strip. In these experiments, an 820 nm-wavelength laser beam is used to excite photoconductive Auston switch, while the magnetization is extracted from the Kerr signal of a reflected 410 nm-wavelength laser beam. In both instances, the pulse duration is between 150-200 fs.

The dynamic TR-MOKE measurements are achieved by using a pulsed Ti:sapphire laser with a central wavelength of 800 nm, a pulse duration of 150 fs, and a repetition rate of 1000 Hz. An intense pump pulse beam with a fluence of approximately 1.0 mJ/cm$^2$ is used to excite the dynamic magnetization behaviors, and the transient MOKE signal is detected by a weak probe pulse of approximately 0.05 mJ/cm$^2$, which is time delayed with respect to the pump beam. The pump and probe laser beams are at almost perpendicular incidence, with spot diameters of about 1.0 and 0.2 mm, respectively. During the TR-MOKE measurement, an external field $H$ is applied in the *x-z* plane, with a polar angle of $\theta_H$=71° to drive the magnetization orientation away from the perpendicular easy axis.

**Acknowledgements**

This work is supported by the NSF Award No. 1611570, the Nanosystems Engineering Research Center for Translational Applications of Nanoscale Multiferroic Systems (TANMS), and the Spins and Heat in Nanoscale Electronic Systems (SHINES), an Energy Frontier Research Center funded by the US Department of Energy (DOE). The authors are also grateful for the support from the Function Accelerated nanomaterial Engineering (FAME) Center, and a Semiconductor Research Corporation (SRC) program sponsored by Microelectronics Advanced Research Corporation (MARCO) and Defense Advanced Research Projects Agency (DARPA). G.

Wu and Z. Zhang are supported by the National Natural Science Foundation of China (Grant Nos. 11874120 and 51671057).

**References:**

1. Myers, E. B., Ralph, D. C., Katine, J. A., Louie, R. N. & Buhrman, R. A. Current-Induced Switching of Domains in Magnetic Multilayer Devices. *Science* **285**, 867-870, doi:10.1126/science.285.5429.867 (1999).
2. Slonczewski, J. C. Current-driven excitation of magnetic multilayers. *Journal of Magnetism and Magnetic Materials* **159**, L1-L7, doi:https://doi.org/10.1016/0304-8853(96)00062-5 (1996).
3. Wang, K. L., Wu, H., Razavi, S. A. & Shao, Q. in *2018 IEEE International Electron Devices Meeting (IEDM)*. 36.32.31-36.32.34.
4. Maruyama, T. *et al.* Large voltage-induced magnetic anisotropy change in a few atomic layers of iron. *Nature Nanotechnology* **4**, 158-161, doi:10.1038/nnano.2008.406 (2009).
5. Li, X., Lee, A., Razavi, S. A., Wu, H. & Wang, K. L. Voltage-controlled magnetoelectric memory and logic devices. *MRS Bulletin* **43**, 970-977, doi:10.1557/mrs.2018.298 (2018).
6. Fan, Y. *et al.* Magnetization switching through giant spin–orbit torque in a magnetically doped topological insulator heterostructure. *Nature Materials* **13**, 699-704, doi:10.1038/nmat3973 (2014).
7. Han, J. *et al.* Room-Temperature Spin-Orbit Torque Switching Induced by a Topological Insulator. *Physical Review Letters* **119**, 077702, doi:10.1103/PhysRevLett.119.077702 (2017).
8. Dc, M. *et al.* Room-temperature high spin–orbit torque due to quantum confinement in sputtered $Bi_xSe_{(1-x)}$ films. *Nature Materials* **17**, 800-807, doi:10.1038/s41563-018-0136-z (2018).
9. Khang, N. H. D., Ueda, Y. & Hai, P. N. A conductive topological insulator with large spin Hall effect for ultralow power spin–orbit torque switching. *Nature Materials* **17**, 808-813, doi:10.1038/s41563-018-0137-y (2018).
10. Wu, H. *et al.* Room-Temperature Spin-Orbit Torque from Topological Surface States. *Physical Review Letters* **123**, 207205, doi:10.1103/PhysRevLett.123.207205 (2019).
11. Wu, H. *et al.* Spin-Orbit Torque Switching of a Nearly Compensated Ferrimagnet by Topological Surface States. *Advanced Materials* **31**, 1901681, doi:10.1002/adma.201901681 (2019).
12. Stanciu, C. D. *et al.* Ultrafast spin dynamics across compensation points in ferrimagnetic GdFeCo: The role of angular momentum compensation. *Physical Review B* **73**, 220402, doi:10.1103/PhysRevB.73.220402 (2006).
13. Kirilyuk, A., Kimel, A. V. & Rasing, T. Ultrafast optical manipulation of magnetic order. *Reviews of Modern Physics* **82**, 2731-2784, doi:10.1103/RevModPhys.82.2731 (2010).
14. Stanciu, C. D. *et al.* All-Optical Magnetic Recording with Circularly Polarized Light. *Physical Review Letters* **99**, 047601, doi:10.1103/PhysRevLett.99.047601 (2007).
15. Kimel, A. V. *et al.* Ultrafast non-thermal control of magnetization by instantaneous photomagnetic pulses. *Nature* **435**, 655-657, doi:10.1038/nature03564 (2005).
16. Lambert, C.-H. *et al.* All-optical control of ferromagnetic thin films and nanostructures. *Science* **345**, 1337-1340, doi:10.1126/science.1253493 (2014).
17. Mangin, S. *et al.* Engineered materials for all-optical helicity-dependent magnetic switching. *Nature Materials* **13**, 286-292, doi:10.1038/nmat3864 (2014).
18. Xu, Y. *et al.* Ultrafast Magnetization Manipulation Using Single Femtosecond Light and Hot-Electron Pulses. *Advanced Materials* **29**, 1703474, doi:doi:10.1002/adma.201703474 (2017).
19. Yang, Y. *et al.* Ultrafast magnetization reversal by picosecond electrical pulses. *Science Advances* **3**, e1603117, doi:10.1126/sciadv.1603117 (2017).
20. Radu, I. *et al.* Transient ferromagnetic-like state mediating ultrafast reversal of antiferromagnetically coupled spins. *Nature* **472**, 205-208, doi:10.1038/nature09901 (2011).

21   Smith, F. W. *et al.* Picosecond GaAs-based photoconductive optoelectronic detectors. *Applied Physics Letters* **54**, 890-892, doi:10.1063/1.100800 (1989).
22   Miron, I. M. *et al.* Perpendicular switching of a single ferromagnetic layer induced by in-plane current injection. *Nature* **476**, 189-193, doi:10.1038/nature10309 (2011).
23   Liu, L., Lee, O. J., Gudmundsen, T. J., Ralph, D. C. & Buhrman, R. A. Current-Induced Switching of Perpendicularly Magnetized Magnetic Layers Using Spin Torque from the Spin Hall Effect. *Physical Review Letters* **109**, 096602, doi:10.1103/PhysRevLett.109.096602 (2012).
24   Liu, L. *et al.* Spin-Torque Switching with the Giant Spin Hall Effect of Tantalum. *Science* **336**, 555-558, doi:10.1126/science.1218197 (2012).
25   Roschewsky, N., Lambert, C.-H. & Salahuddin, S. Spin-orbit torque switching of ultralarge-thickness ferrimagnetic GdFeCo. *Physical Review B* **96**, 064406, doi:10.1103/PhysRevB.96.064406 (2017).
26   Ju, G. *et al.* Ultrafast Time Resolved Photoinduced Magnetization Rotation in a Ferromagnetic/Antiferromagnetic Exchange Coupled System. *Physical Review Letters* **82**, 3705-3708, doi:10.1103/PhysRevLett.82.3705 (1999).
27   van Kampen, M. *et al.* All-Optical Probe of Coherent Spin Waves. *Physical Review Letters* **88**, 227201, doi:10.1103/PhysRevLett.88.227201 (2002).
28   Mekonnen, A. *et al.* Femtosecond Laser Excitation of Spin Resonances in Amorphous Ferrimagnetic $Gd_{1-x}Co_x$ Alloys. *Physical Review Letters* **107**, 117202, doi:10.1103/PhysRevLett.107.117202 (2011).
29   Wu, G., Chen, S., Ren, Y., Jin, Q. Y. & Zhang, Z. Laser-Induced Magnetization Dynamics in Interlayer-Coupled [Ni/Co]$_4$/Ru/[Co/Ni]$_3$ Perpendicular Magnetic Films for Information Storage. *ACS Applied Nano Materials* **2**, 5140-5148, doi:10.1021/acsanm.9b01028 (2019).